# Electric Stimulation of the Retina

by
Erich W. Schmid[1] and Robert Wilke[2]

[1] Institute of Theoretical Physics, Tuebingen University, Auf der Morgenstelle 14, 72076 Tuebingen, Germany
[2] Graduate School of Biomedical Engineering, University of New South Wales, Sydney, NSW 2052, Australia

**Abstract:**
Two computational models to be used as tools for experimental research on the retinal implant are presented. In the first model, the electric field produced by a multi-electrode array in a uniform retina is calculated. In the second model, the depolarization of the cell membrane of a probe cylinder is calculated. It is shown how these models can be used to answer questions as to cross talk of activated electrodes, bunching of field lines in monopole and dipole activation, sequential stimulation, etc. The depolarization as a function of time indicates that shorter signals stimulate better, as long as the current does not change sign during stimulation.

## 1. Introduction

Since more than 15 years researchers are trying to restore part of the vision of people suffering from retinitis pigmentosa by implanting a multi-electrode array in subretinal position or in epiretinal position [Ch 1993, Ch 2004, Pa 2005, Mc 2007, Ho 2008, Hu 2009, Zr 2010]. A pixelized image is transmitted to the human retina and into the visual pathway via electric stimulation. Early designs using photodiodes as power source failed for lack of power. Later designs with energy supplied from external power sources report remarkable success [Hu 2009, Zr 2010]. Despite success results are not perfect and there is still need for improvement. The desire is that a blind person with retinal vision prosthesis can notice obstacles while walking, or can recognize persons from seeing their face.

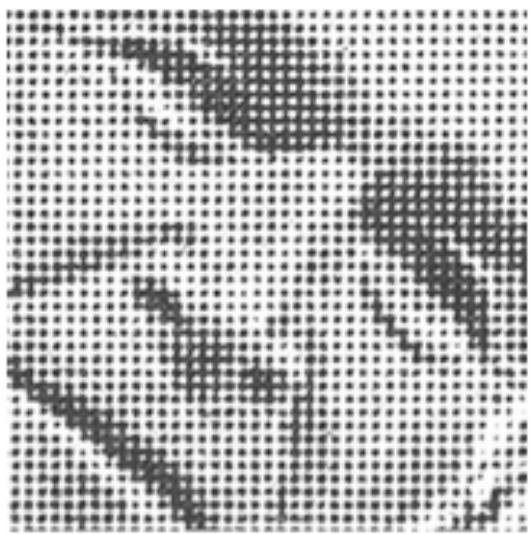

Fig. 1.1. Image section with 1500 pixels, copied from a newspaper.

Let us see what is needed to recognize a face. Fig. 1.1 shows an image section taken from a group photo in the local newspaper. It has about 1500 pixels. Viewed from the right distance one clearly recognizes a former president of the United States. This is not trivial, because one does not recognize the person, for instance, when the viewing distance is too close. And it is not only the viewing distance; there are more prerequisites that must be met before one can recognize the person: 1. Every pixel must be confined to its location on the mesh, without overlap with its neighbor. 2. Pixels must be tunable independently from each other. 3. The grey value of every pixel must be tunable from dark to bright with accuracy in the percent range.

Can these prerequisites be met by electric stimulation of the retina? We have to expect problems. Phosphenes created by electric stimulation may be bigger than the area allowed for one pixel. Neighboring phosphenes might influence each other via cross talk. It might be difficult to tune the brightness of pixels with the needed accuracy without increasing their size.

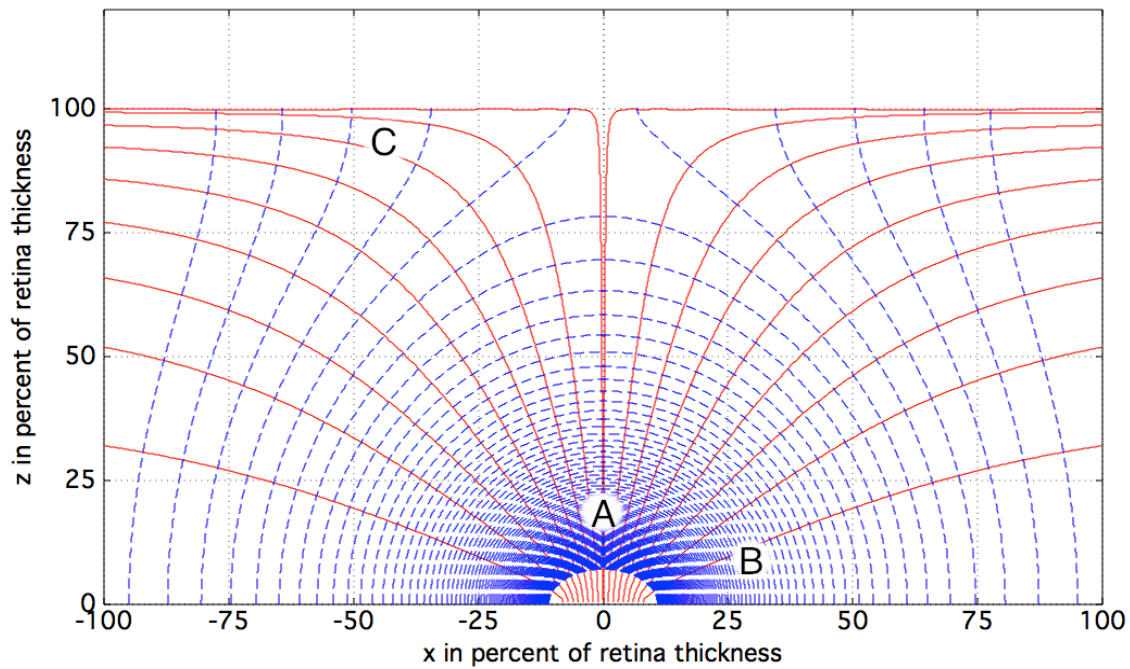

Fig. 1.2. One electrode in the center of a subretinal multi-electrode array is activated. The vitreous has been replaced by oil of extremely small electric conductivity. Current lines are shown in red (full lines), equipotential lines in blue (broken lines). Areas where the electric signal might possibly enter into the visual pathway are indicated by A, B and C (see text).

To date, retinal prostheses are still far away from producing a percept that is comparable in quality with the image shown in Fig. 1.1. More research and novel ideas will be needed before we reach this goal. In order to see where we are with our present knowledge, let us look at a picture that will be explained in more detail in Sect. 2: Fig. 1.2 shows the electric current lines (full lines, red) emerging from a single electrode placed in subretinal position. Equipotential lines are also shown (broken lines, blue). The vitreous has been removed and replaced by silicone oil with negligible electric conductivity. The patient sees a yellowish white phosphene. The question is: where does the electric signal enter into the visual pathway? The possibilities are:

> Area A. The electric field is almost vertical and rather strong, the volume of the area is small. The prime targets are bipolar cells. One expects a low threshold and small phosphenes.
>
> Area B. The electric field is almost horizontal. With monopole excitation, this area has the shape of a torus and, therefore, its volume is a little bigger than area A. Area B is ideal for electrodes operating as dipoles.
>
> Area C. The electric field is almost horizontal and weak. The area is widespread. One expects interference of the electric signal with neural network processing in the ganglion cell layer and inner plexiform layer. Phosphenes are expected to have a size similar to the thickness of the retina (when their angular size is visualized by projecting from the plane of perception back to the retina).

Many experiments have been performed in vitro in order to find out details of electric stimulation. Everyone of the areas A,B,C has a signature with respect to experimental parameters, such as threshold, blocking by chemicals, etc. Despite great efforts there are still many open questions. For the design of a retinal prosthesis the interesting quality of a percept is its brightness and size, which are hard to measure in vitro or by animal experiments. A few measurements have been done with human test subjects [Wi 2006, Mc 2007, Ya 2007, De 2008, Ho 2008, Na 2008, Wi 2008, Ca 2009, Gr 2009, Na 2009, Wi 2009, Wi(1) 2010, Wi(2) 2010]. Rather large phosphenes have been found in these investigations. Does that mean that stimulation takes place in area C? We don't know the answer, but we don't want to exclude this possibility. All three possibilities need further investigations, both experimental and theoretical.

In this paper we want to present theoretical tools, which will be helpful in finding answers, either by proposing experiments, or by analyzing experimental results. Simple theoretical models, simple algorithms and their coded implementations will be presented, tested and discussed. The techniques are simple, seen from the level of modern computational physics. Emphasis is on application to problems of electric stimulation and imaging. Examples will be given for the subretinal position of the multi-electrode array. The computer codes will work equally well for an epiretinal position of the array.

In Sect. 2 quasi-static electric fields generated by multi-electrode arrays in a model tissue will be calculated. A neuron and its processes can detect the electric field associated with a current in a biological tissue. The field enters into the nerve cell much in the way the field of an electric signal in the atmosphere enters into an antenna. The equation that governs this process is called cable equation. Sect. 3 will present the needed simplifications, the cable equation, a method of solution and examples. A summary will be given and conclusions will be drawn in Sect. 4.

## 2. Calculation of the electric field

In order to study electric stimulation we need to know the electric field that stimulates. We will not be able to calculate it for realistic electrodes and for the full complexity of the retinal tissue. What can be done, however, is to define a continuum model and to do the calculation within the limits of this model. In the next sections we will present such a model, discuss the equation to be solved, the computer method used and some typical results.

### 2.1. Continuum approximation of the retinal tissue

The retinal tissue consists of various types of cells, with rather small clefts in between. The quasi-static electric current that we are considering in this section has to follow these extracellular clefts. We call it an Ohmian current, in contrast to a displacement current that will be considered in a future paper.

In the continuum approximation cells and clefts form an electric conductor with an average electric conductivity $\sigma$. There is an electric current density *j(x,y,z,t)* and an electric field *E(x,y,z,t)*. The two are related by Ohms' law

$$j(x,y,z,t) = \sigma E(x,y,z,t). \qquad (2.1)$$

In a refined model, the electric conductivity would be a tensor, with components depending on the space coordinates *x,y,z*. Unfortunately, very little information is available on the components of this tensor in case of the human retina. Therefore, and for reasons of simplicity of our computer programs, we replace the tensor by a scalar, as is seen in Eq. (2.1). We allow, however, one refinement: there may be two horizontal layers with two different values of σ. This will allow us to study the effect of a thin layer of extracellular fluid between the multi-electrode array and the retina. It will also allow us to calculate the current field when the vitreous is replaced by silicone oil. The result of such a calculation has already been shown in Fig. (1.2).

## 2.2. The multi-electrode array

In our calculations the multi-electrode array will be flat, with electrodes that are either flat or protruding like small hemispheres. No conical electrodes or needle electrodes will be considered. As the arrangement of electrodes on the chip we consider the quadratic Bravais lattice and the hexagonal Bravais lattice; such lattices are well known from chessboard and honeycomb structures.

In most calculations the counter electrodes will be located on the array. In fact, every electrode may become a counter electrode by changing the applied voltage. When multipoles formed by electrodes are not charge-balanced a counter electrode at infinity will appear automatically. Whether all counter electrodes are on the chip or part of the electric current (or all of it) has to go to infinity will have severe consequences on the electric stimulation.

## 2.3. The Poisson equation

The equation to be solved is called the Poisson equation

$$\nabla \cdot \vec{j}(x,y,z,t) = -\nabla \cdot \left[\left(\sigma + \varepsilon_0 \varepsilon_r \frac{\partial}{\partial t}\right) \nabla V(x,y,z,t)\right] = 0 \qquad (2.2)$$

Nabla-dot means *divergence,* nabla without dot means *gradient. V(x,y,z,t)* is the potential field that drives the current *j(x,y,z,t)*. The *gradient* of the potential field is the electric field *E(x,y,z,t)*. The current has an Ohmian part and a displacement part. The latter arises from the operator with the time derivative. Since we are considering the quasi-static case, this part can be dropped. Also, there is no need in our quasi-static case to carry along the time coordinate *t*. The electric conductivity σ is a non-zero constant. Let us postpone the treatment of two layers with two different values for σ. If there is only one layer we can divide the equation by σ and arrive at the Laplace equation

$$\nabla \cdot \nabla V(x,y,z) = \Delta V(x,y,z) = 0. \qquad (2.3)$$

The equation is valid in a volume without current sources or sinks.

## 2.4. The boundary condition

The solution of the Laplace equation (2.3) is unique in a closed space when the potential function *V(x,y,z)* at the surface of that space is given. The closed space may be simply connected or non-simply connected.

Our present problem is how to let the electrodes enter into the calculation. They have to enter via the prescription of an adequate boundary condition. The surface of the electrodes, i.e. the electrode-tissue interface, is a highly complicated system, as we can learn from voltammetry. For realistic electrodes, there is a problem of time-dependence, even in the case when a con-stant voltage is applied to the electrode, and even in the case of current control. This comes from the fact that corners and edges of the electrode fire first, while the center of the electrode emits its current later. At present such details are not in the focus of our interest.

In order to stay clear of such complications, we put the electrodes into little capsules and prescribe the boundary condition at the surface of these capsules. The capsules are supposed to be just a little bigger than the electrodes, but big enough to avoid the region of complication. If the shape of the electrode (circular or square, for instance) plays a role the capsule might be a small cylinder or a box. If the target volume in which we want to study electric stimulation is a little farther away from the surface of the electrode, the capsule may be spherical. In most of the examples presented in this paper we assume that this is the case.

A layer of small spheres, in our computational model, thus represents the multi-electrode array. This layer forms the x,y-plane of the coordinate system, with origin at the center of the array, and small hemispheres sticking out above and below the x,y-plane. The z-axis points into the retina, from the assumed subretinal or epiretinal position of the electrode array.

So far, the space in which we want to solve the Laplace equation is still open. We close it by a large sphere around the origin of the x,y,z coordinate system, with radius going to infinity. We do not take a hemisphere because we do not want the x,y-plane to become a boundary. We want the x,y-plane to become an element of mirror symmetry, and disregard the solution for negative values of z.

We are now ready to write down the boundary condition. On the surface of the large sphere we demand that the potential is zero. On the surface of the little capsules we have to generate the boundary potential by virtual charges inside the capsules. These charges are either sitting on the x,y-plane, or form pairs of mirror symmetry below and above the x,y-plane. Whenever the target volume, i.e. the volume in which we expect the stimulation signal to excite a neuron, is more than one or two electrode diameters away from the capsule, we will need only one virtual charge in the center of the capsule.

These virtual charges are not to be mistaken with physical charges on electrodes. The only purpose of virtual charges is to create a boundary potential at the surface of the capsule, nothing more.

## 2.5. The method of solution

We are using a method described by E. Kasper [Ka 2001]. This method reduces the solution of the Laplace equation to a variational problem, namely the problem of optimizing the values of the virtual charges.

The virtual charges inside the capsules are chosen, independently from each other, with the goal of reproducing the prescribed boundary condition at the surfaces of the capsules. One should note, however, that the potential at the surface of a capsule is the sum of the Coulomb potentials of *all* virtual charges, not only of the virtual charges inside one capsule. Already here one can see that we have to expect cross talk between capsules, even long-ranged cross talk, provided that the firing of electrodes is unipolar and simultaneous.

Before giving numerical values to the virtual charges we have to distinguish between two kinds of stimulation signals:

1. The *voltage-controlled* stimulation signal. In this case a voltage generator applies a certain voltage to an electrode (a square pulse, a ramp, or other) and thus induces a current in the tissue. This current does not only depend on the applied voltage but also on the (time-dependent) amount of voltage drop in the Helmholtz layer at the surface of the electrode; a typical case will be seen in Fig. 3.3b. Fortunately, the plot of electric field lines will be scale invariant. This fact allows us to pick just one moment of the time dependent profile of the signal for setting the boundary condition. In all examples shown the condition will be a prescribed value of the potential function at the zenith of the little hemisphere. This point lies inside the tissue, shortly above the Helmholtz layer. When more than one electrode fire simultaneously there will be more such points with prescribed potentials, and there will be more virtual charges to be chosen. They will be chosen by variation, either by hand or by the Levenberg-Marquardt method.

2. The current controlled stimulation signal. In this case a current generator adjusts the voltage applied to an electrode such that the current injected into the tissue has a prescribed value (constant in time, or of prescribed time-dependence). For our computer code this will be the more simple case, because no variation of virtual charges is needed. The Gauss-Ostrogradsky theorem tells us that the value of the surface-integral over the electric field component perpendicular to the surface of the capsule, after multiplication with σ, gives us the current. In simple words: the current injected into the tissue and the virtual charge inside the capsule differ only by a constant factor.

Let us recollect what we have so far. There is the large volume of a sphere with radius going to infinity. In this volume, with exclusion of little capsules, we want to calculate the solution $V(x,y,z)$ of the Laplace equation. And we have virtual charges inside the capsules, with values determined by variation, or determined by the currents injected by a current generator. As soon as we know the position and values of the virtual charges it is easy to obtain the solution $V(x,y,z)$: it is simply the sum of the Coulomb potentials of the virtual charges!

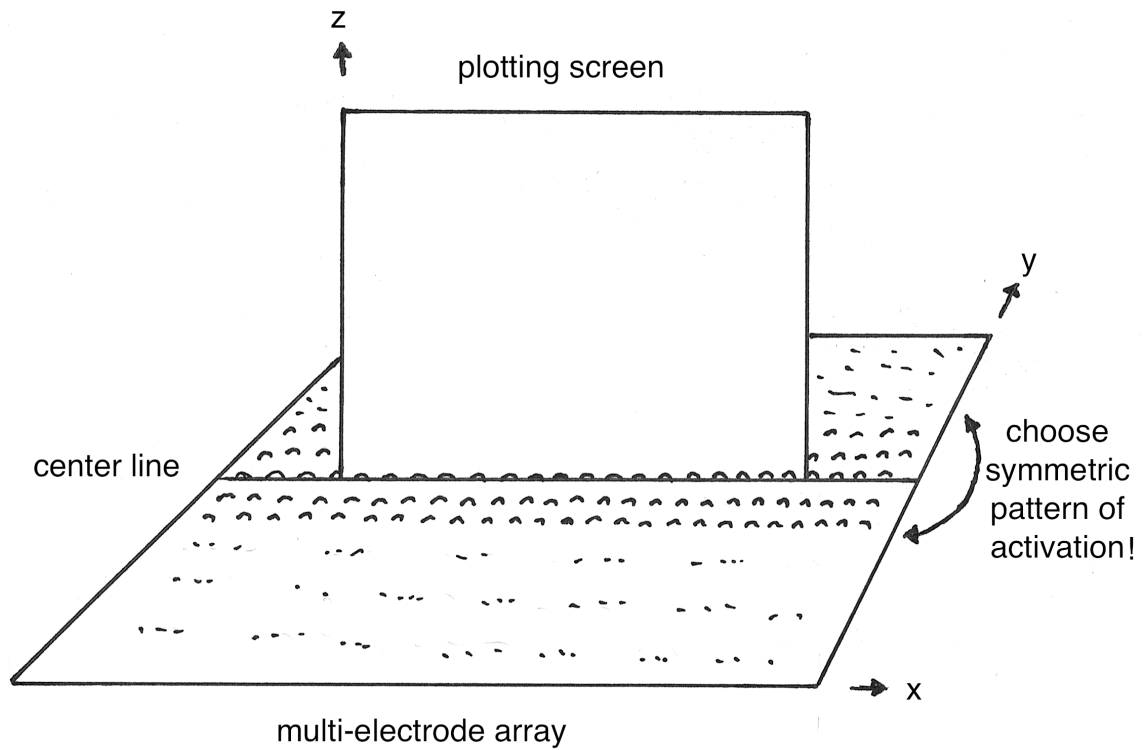

Fig. 2.1. The electrode array and the plot screen.

Actually we don't want to know the solution *V(x,y,z)* everywhere. We are not interested in *V(x,y,z)* for negative values of *z*, as has been said before. The region of the lower big hemisphere is unphysical. It was only necessary because we did not want the x,y-plane to be a boundary. Now we just ignore that lower hemisphere.

The chosen plot screen is seen in Fig. 2.1. The field lines must lie on the screen and are not allowed to penetrate the screen. For this reason we have to impose a symmetry condition. We demand that the pattern of electrode activation for positive values of y and for negative values of y is the same. This condition guarantees that the vector of the electric field (as well as the vector of the electric current) does not have a component perpendicular to the plot screen.

Things are more complicated when the medium for which we are calculating currents and fields consists of two horizontal layers having different electric conductivity. We want to be able to study two such cases: (1) an adhesive layer of saline between the chip and the retina, and (2) a vitreous replaced by silicone oil (as is seen in Fig. 1.2.). For solving the two-layer problem we employ the method of multiple image charges, as described by Lehner [Le 1990]. In this method every one of the virtual charges becomes the starting point of an infinite, but fast-converging, series of charges. We skip the details and refer to the literature.

## 2.6 Examples and discussion of results

Fig. 2.2 shows two extreme cases. In Fig. 2.2a only the center electrode of the electrode array is activated. There is exactly one virtual charge at the center of the hemisphere. It produces as boundary condition a value of 1 Volt at all points on the hemisphere. The full (red) lines show the direction of the electric field or, via Eq. (2.1), the direction of the electric current. The broken (blue) lines are the equipotential lines. They start at +1 V at the hemisphere and have a spacing of 0.1 V. The electric field has radial direction. The potential decreases as 1/r (with r being the radial distance from the electrode), as expected.

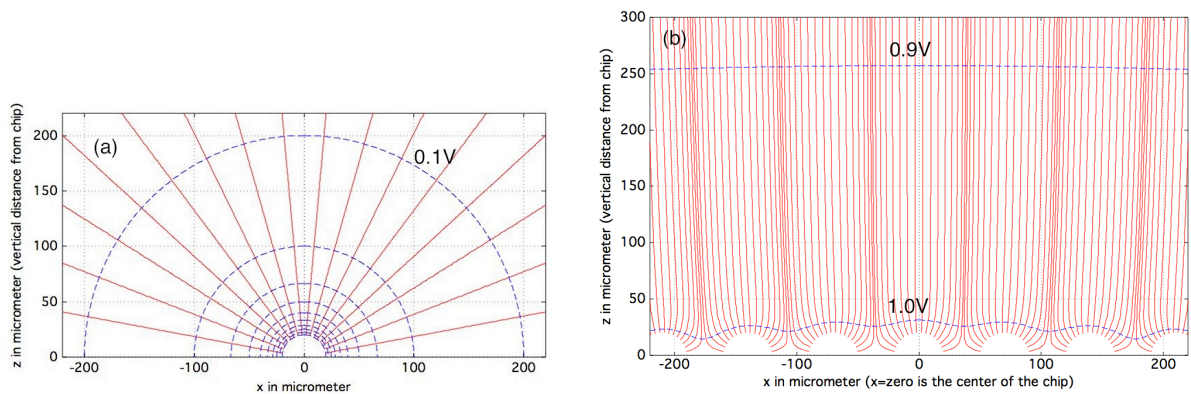

Fig. 2.2. Electrical field lines (or current lines) are shown as full (red) lines, equipotential lines as broken (blue) lines. (a) Only the center electrode of the array is activated. (b) All 1521 electrodes are activated, simultaneously and unipolar.

In Fig. 2.2b all electrodes, namely 39 rows by 39 columns, in a cubic Bravais lattice of electrodes, are activated. For the blind person this means a "structure less white wall". All of the 1521 electrodes enter into the calculation. A section of the full plot is displayed. Again, there is only one virtual charge at the centers of the 1521 hemispheres. Their values have been determined by a 3-parameter variation with a 1 % tolerance in reproducing the boundary potential of 1 Volt at the summit of each little hemisphere. The lower broken line is the 1 Volt equipotential line, as in Fig. 2.2a. It does not touch the hemisphere, but lies within the tolerance, as it should. The spacing of the equipotential lines is 0.1 Volt, as in Fig. 2.2a. The electric field lines (current lines) have become almost vertical and the strength of the electric field (or current) has become rather small, probably too small for electric stimulation, even if the voltage applied to the electrodes is doubled!

Fig. 2.2a is instructive, in a certain sense. It tells us that we cannot reproduce the strength of the electric field (or current) by the density of plotted lines. Their density goes down with $1/r$ for the full (red) lines, while the strength of the field (or current) goes down with $1/r^2$. The $r^2$, instead of $r$, comes from the fact that current travels in a cone and not in a two-dimensional stripe, as seen on the plot. We don't need to show the strength of the electric field (or current) by the density of lines; we get the strength of the field by looking at the spacing of the broken (blue) equipotential lines.

The calculation behind Fig. 2.2b demonstrates the amount of cross talk we are obtaining when many electrodes fire simultaneously and unipolar. The result of the variational calculation for the values of the virtual charges tells us this: The hemisphere in the middle of Fig. 2.2b needs only about 2 % of the virtual charge needed in case of Fig. 2.2a to produce the +1 V at the pole of the hemisphere. The rest of 98% comes by cross talk from the other electrodes!

One might think that a current-controlled stimulation signal could be a way out of the dilemma. Well, it is very simple to let the same amount of current that is used in Fig. 2.2a emerge from every one of the 1521 electrodes. We performed the calculation and obtained a plot that looks similar to the one shown in Fig. 2.2b; we find new equipotential lines in the vicinity of the old ones. Only that the new lower equipotential line now has a value of about 39 Volt and the new upper equipotential line has a value of 33 Volt. Such voltages seem to be too high to be applied in a biological system.

Disregarding the problem of too weak or of too strong electrical fields we tried to perform some imaging. As has been said before, all electrodes activated simultaneously means imaging a "white wall" for the blind person carrying the retinal prosthesis. If we turn off the center electrode, the white wall should get a black spot in the middle. If we turn off the center column of electrodes, we should see a vertical black line on a white wall. And if we turn off every column of electrodes with odd column-index, we should see black and white stripes. We performed these calculations and found out that the electric field in fact does not show any of these structures, except at extremely short distances from the electrodes, i.e. at distances below the lattice constant. The cross talk washed out all these structures. To be precise: the computer can see the small modifications of the field. And a precise instrument would also be able to detect them, but not likely a human subject with a retinal prosthesis.

In order to understand the physics behind these findings it is helpful to go to a somewhat extreme case. We perform a calculation for over ten thousand electrodes, activated under current control. Every one of the many electrodes emits the same current as the electrode in Fig. 2.2a. The full plot is shown in Fig. 2.3. The lowest equipotential line (broken, blue) now has a value of 104 Volt. The spacing of equipotential lines in the plot is 1 Volt.

What is seen in Fig. 2.3 can be compared with textbook knowledge. In the center part, e.g. from x = -500 μm to +500 μm and z = 0 to 500 μm, we see approximately the homogeneous field of a plate capacitor, with a somewhat rough lower plate and an upper plate far away. At the end of the chip, near x = 3500 μm, we see the edge effect: a maximum of electric field strength.

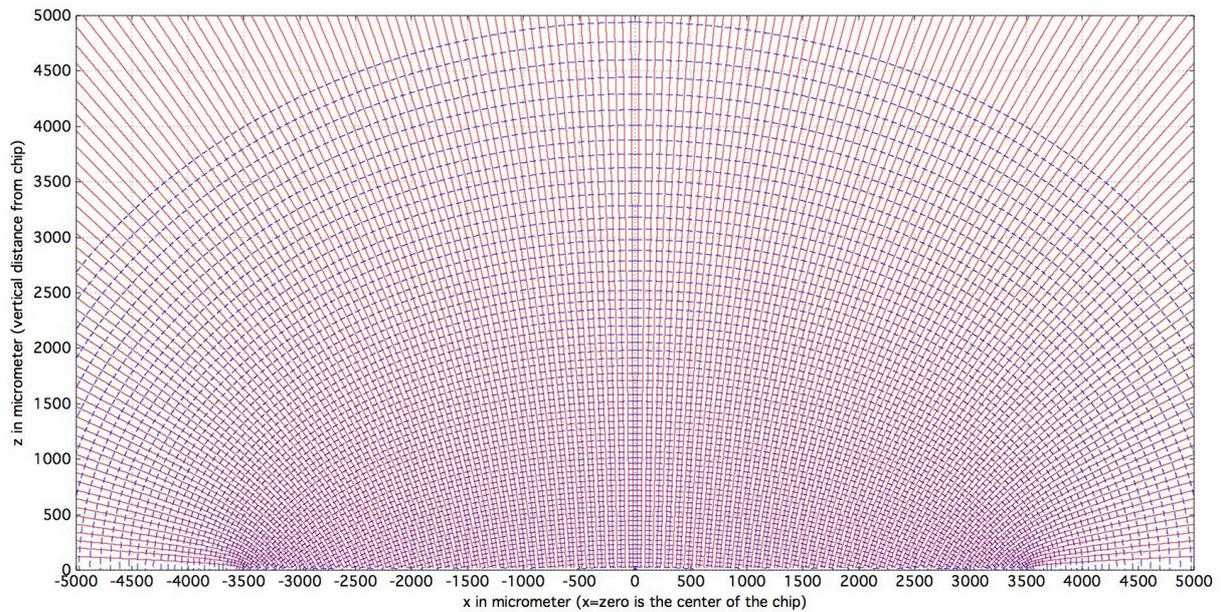

Fig. 2.3. Simultaneous and unipolar activation of 10201 electrodes (101 rows by 101 columns). Every electrode injects the same current as the one electrode in Fig. 2.2a. The spacing of electrodes is 70 μm in both x- and y-direction. Electric field lines are shown in red (full lines), equipotential lines are shown in blue (broken lines). The equipotential lines start at 104 Volt at the bottom, near x=0, and have a spacing of 1 Volt.

Let us get a simple physical explanation for the high voltage of 104 Volt at the bottom, near x=0. In Fig. 2.3 every electrode sends out the same electric current, as does the one electrode in Fig. 2.2a. This current is represented by exactly one red (full) line, in the plot. Between every two of these lines there exists a separatrix, which means a dividing line of two current fields that have different sources, or sinks, and cannot interpenetrate. In three dimensions, these separatrices form hoses that separate the currents of the various electrodes. The hoses are narrow at their beginning and open up like trumpets towards infinity; there is only a fraction of the full solid angle, for every one of these trumpets. The more hoses there are, the narrower they get, and the higher their electrical resistance gets! It is this latter fact that increases the voltage drop for given current or decreases the current, for given voltage drop. If the vitreous is replaced by silicone oil the squeezing of field lines becomes more severe, with one exception: there is a stagnation point in the center of the chip with no current at all.

We can also see how the method allows us to model a single electrode. In Fig. 2.3, what we get by zooming out is the shape of a field that would be obtained by using a finite element method for a single electrode, with uniform emission of current from all elements. We shall use this possibility a little later.

From Fig. 2.3 we can also learn something about imaging. One cannot present an image without a sharp contrast between a white and a black part of the image. In Fig. 2.3 the uniformly activated chip represents the image of a "white wall". The surrounding of the chip represents a black background. How much contrast do we see between the two? At the border between white and black, at 3500 μm, the electric field is very strong, as has been said before. This strong field surely becomes weaker if we go out further. But it does not decay fast enough to give the desired contrast. In fact, we don't see anything that looks like a sharp contrast. Fortunately this result is unimportant, because we cannot allow a voltage of 100 Volt anyway for biological reasons.

The results we have seen so far are telling us that simultaneous unipolar stimulation is not suitable for imaging, because of cross talk between electrodes. This has been said by Daniel Palanker [Pa 2005] and by one of the authors of this manuscript [Sc 2005].

There are two physical ways to avoid the cross talk problem. The first one is to give up simultaneous stimulation by going over to sequential activation of electrodes. The second way is to give up unipolar stimulation by using multipoles. There are some obstacles for both of these ways that can be further studied in computational models.

Sequential stimulation is a matter of counting time slices, in the first place. We need about twenty frames per second to present a movie to the blind person. That means 50 milliseconds per frame. Into how many time slices can we divide such a frame? In a voltage controlled 500-microsecond pulse it takes about 1 millisecond for charging and discharging the electrode. Shorter pulse lengths are under discussion, especially asymmetric ones with fast charging of electrodes, and slow subthreshold discharging of electrodes [Sc 2007]. Shorter pulse length increases the number of time slices. But let us stick to time slices of 1 millisecond, just for simplicity of the discussion. This then means 50 time slices per image frame. But we are also discussing 1521 electrodes. On a bright day, or in front of a white wall, most of these electrodes will fire. This means that around 30 electrodes will fire in one time slice simultaneously. Let us do some model calculations.

We start with the following situation. There is a dark room and a table with a black tablecloth. On the tablecloth there is a small bright object. There is no cross talk, or almost no cross talk. The calculation for this case has already been done. The result is seen in Fig. 2a. Now we turn the lights on and use a white tablecloth. For simultaneous stimulation, the result is seen in Fig. 2b. But now we are passing to sequential stimulation. How much distortion of the electric field in the middle of the chip will we get? We did the calculation for two cases. In the first case 24 electrodes, evenly spaced over the chip, are engaging in cross talk with the electrode in the middle. In the second case, 8 electrodes, sitting in the 4 corners of the chip and at the ends of the x- and y-axis are firing and disturbing by cross talk the field of the center electrode. The result is shown in Fig. 2.4a,b. We see the distorted current lines and equipotential lines and have to compare them with those of Fig. 2.2a.

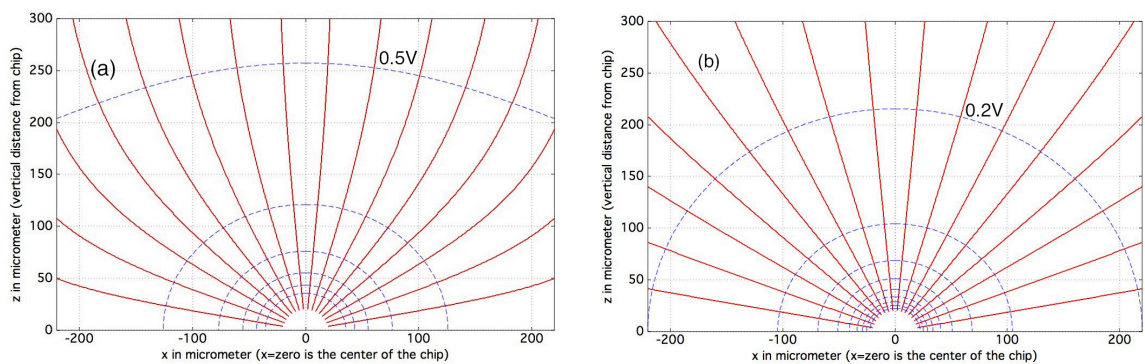

Fig. 2.4. The electric field of the center electrode is shown under the influence of cross talk (undistorted, this field should be identical to the one of Fig. 2.2a). In (a) the center electrode is under the influence of cross talk coming from 24 activated electrodes, evenly spread over the array. In (b), 8 electrodes spread around the rim of the array are distorting the field.

With 24 activated electrodes, in addition to the electrode at the center, the distortion by cross talk is rather strong. At a penetration depth of 200 μm into the retina the potential of the stimulating current has become more than five times stronger. With only 8 additional activated electrodes the potential is still twice as strong as without cross talk. The present model is too crude to draw precise conclusions on imaging. It may still be possible to perform some basic imaging with 8 cross-talking electrodes, but it is rather unlikely that the fine-tuning needed to transmit grey values will be possible in this case.

The second way of avoiding cross talk is to give up unipolar stimulation by using multipoles, as has been said before. Let us consider in the following pixels formed by dipoles. Here we are expecting two problems. The first problem arises from the big fraction of the current that travels too close to the surface of the chip and thus is lost for stimulation. The second problem comes from the fast decrease of the strength of a dipole field; while the strength of the electric field of a monopole decreases asymptotically with $1/r^2$, the field of a dipole decreases asymptotically with $1/r^3$.

Let us discuss the first problem in more detail. As we did in examples presented before, we take parameters of the 1521-electrode chip known from the literature [Zr 2010]. Each one of the electrodes has the shape of a square with a side length of 50 μm. Between neighboring electrodes there is a gap of only 20 μm. How much current do we loose through this narrow gap, if two adjacent electrodes form a dipole? To answer this question we performed a model calculation. In this example it is not sufficient to have only one virtual charge in the middle of a hemisphere in order to represent an electrode. We used 169 virtual charges for every one of the two electrodes forming the dipole. This means, we divided each electrode into 13 by 13 little squares. Every square emits the $1/169^{th}$ part of the current that is emitted by the electrode in Fig. 2.2a, and the virtual charges that represent the little squares are chosen accordingly. The result of the calculation is shown in Figs. 2.5a,b. In Fig. 2.5a the retinal tissue is touching the chip. A fraction of 20% of the total current is considered lost because it travels in a layer below 20 μm.

The loss of current increases dramatically if we assume that there is an adhesive saline layer between the chip and the retinal tissue. In the calculation it is assumed that the electric conductivity of the saline is 10 times higher than the one of the tissue. For demonstration purposes we assume that this saline layer is 20 μm thick, which may be exaggerated. The result of the calculation is depicted in Fig. 2.5b. Now 40% of the total current travels in the saline layer below 20 μm and the potential drop inside the tissue that is needed for electric stimulation has become very small.

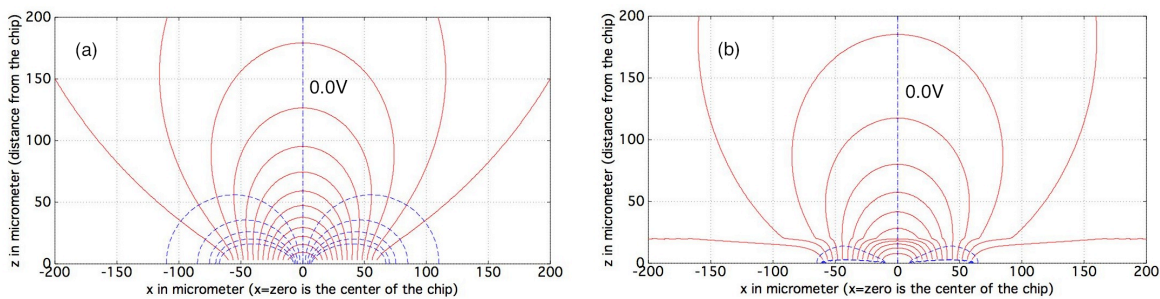

Fig. 2.5. Two adjacent electrodes of the 1521-electrode chip are forming an electric dipole. The electric current injected is equal to the one injected in Fig. 2.2a. The equipotential lines (blue broken lines) have a spacing of 0.1 Volt. (a) The tissue is touching the chip. (b) There is a 20 μm layer of saline with tenfold conductivity between the chip and the tissue.

As a consequence of this calculation we see that one has to form the dipoles with electrodes that lie farther apart from each other. For comparison we stick with the 1521-electrode chip and form a dipole by activating two electrodes while leaving two electrodes in between inactive. The gap between the edges of the two activated electrodes thus becomes 160 μm wide, instead of 20 μm. The result is shown in Fig. 2.6a. For comparison a monopole excitation is shown in Fig. 2.6b. In contrast to Fig. 2.2a we have used 169 virtual charges also for this electrode and we have kept the emitted current fixed. The result is somewhat surprising. The dipole yields a stronger electric field in the region of interest. It is stronger if we are looking for a vertical field in the region "A" of Fig.1.2 where we expect bipolar cells to be the target of stimulation. And we have, in comparison with the monopole, an even stronger horizontal field component in the region "B" of Fig. 1.2 where dendritic connections might be the target.

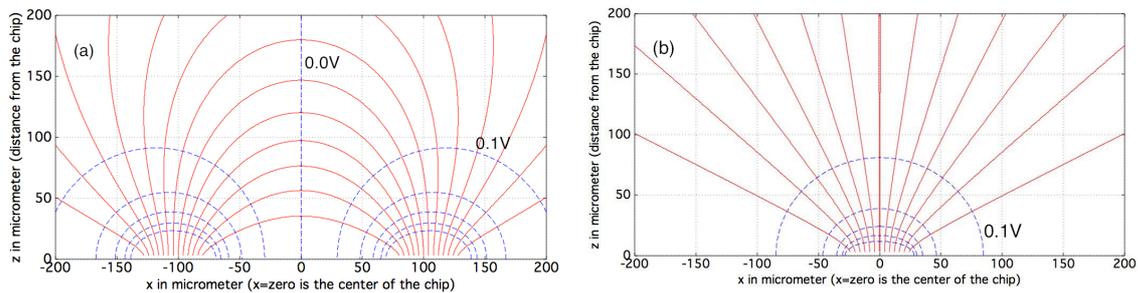

Fig. 2.6. Comparison of dipole and monopole excitation. (a) Two electrodes, one anodic the other one cathodic, emit current into the retina. (b) The same amount of current is emitted by a monopole electrode. The spacing of equipotential lines (blue broken) is 0.1 V, in both plots.

This surprising result merits some physical explanation. It is true that the field strength of an ideal dipole falls off like $1/r^3$. However our dipole is not an ideal one and we are not asymptotically far away. We are in a region where the distance of our target from the two poles is comparable to the distance of the two poles from each other. What is seen in Fig. 2.6a is the bunching effect coming from the second pole. In case of the monopole, Fig. 2.6b, the current is free to spread into a $2\pi$ solid angle. In case of the dipole the current rises into the tissue above an electrode and gets bunched because it has to loop back to the second electrode. Of course, one can compensate for the missing bunching in Fig. 2.6b also by increasing the size of the electrode at the cost of loosing pixel density.

The results of our crude model calculations show that we can indeed avoid long-ranged cross talk by having electrodes and counter electrodes closely (but not too closely) co-located on the same chip. This calls for computer calculations with multipoles: One does not have to stick to dipoles. With tripoles and quadrupoles, for instance, one has the possibility of changing the direction of the horizontal field component [Sc 2005]. It is well known that the retina needs saccades, or at least micro saccades, to avoid image fading. If the area "B" of Fig. 1.2 is the preferred target volume, the variation of field direction could eventually help to substitute for missing micro saccades. Also the bunching effect could be helpful to shape the field of multipoles. More results will be presented in a future paper.

Most of the results presented in this section remain valid also in case of the epiretinal position of the multi-electrode array because only the direction of the z-axis (outward or inward with respect to the center of the eye) is different in the two cases.

# 3. Electric stimulation of neurons and neural networks

The electrical properties of nerve cells have been studied for many years. Starting from the pioneering work by Hodgkin and Huxley [Ho 1952] cellular neurophysiology has rapidly developed and is well described in textbooks [Jo 1995]. Electric stimulation of excitable tissue such as muscle is based on this knowledge and has become a field of research since the first success of pace makers. The amazing success of the cochlea implant has led to an increasing interest in retinal prostheses.

The idea of the subretinal prosthesis is to replace lost rod and cone cells by an array of electrodes. An electrode of such an array sends a weak electric current into the extra-cellular space of the retinal tissue. The electric field associated with this current can enter into the intracellular space of a retinal neuron via the non-zero residual conductivity of the cell membrane and via the membrane capacitance. This so-called passive process can become active in the way described by Hodgkin and Huxley. The stimulation signal carried by the current can thus enter into the visual pathway and lead to the perception of a phosphene. This process is called electric stimulation of the retina. There is hope that it can be used to transmit a pixel image as has been discussed in the Introduction.

The passive stimulation process is described by Heaviside's telegraph equation or more precisely, by the antenna version of the equation. The Heaviside equation describes the physics of a telegraph cable at the bottom of the ocean: The insulation of the cable is good but not perfect, and there is a capacitance that couples the potential inside the cable to the one outside. In case of the ocean cable this coupling causes a gradual loss of signal strength. The opposite process is also possible: the signal travels in the ocean and the cable picks it up like an antenna. Also this antenna process is described by Heaviside's equation; only the driving term of the equation is different.

We want to use Heaviside's cable equation to describe the electric stimulation of neurons. The neuron, or one of its components such as the axon or a dendrite, replaces the cable. The tissue of the retina, in continuum approximation according to the one used in Sect. 2, replaces the ocean. The cell membrane is the insulator of the cable. The antenna process described by Heaviside's cable equation then means that the signal travels in the tissue and enters from there into the neuron or into one of its components.

The signal entering from the extracellular space into the intracellular space will either depolarize or hyperpolarize the negative rest potential of the membrane. The Hodgkin-Huxley model then describes the biological effects that lead from passive depolarization to the active process of signal transmission over longer distances.

The discussion of this Section, like the one of the last Section, is equally valid for epiretinal prostheses and for subretinal prostheses.

## 3.1. The antenna version of Heaviside's cable equation

The cable model is a continuum model. No cell structure is seen except for a section of a neuron (dendrite, axon). A cylinder of length $l$ idealizes this section of a neuron; the cylinder diameter enters into the mathematical treatment only because some constants depend on it.

There is an interior space of the cable and an exterior space. Normally a signal travels in the interior and is attenuated by the coupling to the exterior, which means that it gradually gets lost while traveling. In the case of electric stimulation we have the opposite situation. The stimulation signal travels in the exterior part and enters gradually into the interior part of the

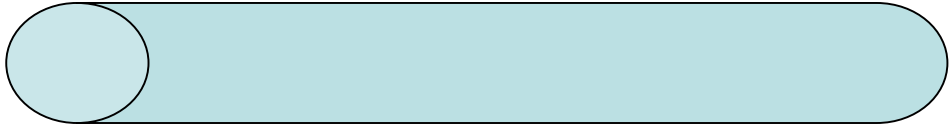

Boundary at *x=0*                                           Boundary at *x=l*

Fig. 3.1. Probe cylinder serving as "antenna" in the antenna version of the Heaviside cable equation.

cylinder. The cable operates like an antenna. The mathematics for both cases is similar, except for an extra driving term in the cable equation.

In our physics model the cylinder is a section of an electric cable. Inside the cable we have the intracellular potential $V_i(x,t)$. Outside the cable we have the extra-cellular potential $V_o(x,t)$. Across the membrane we have the (trans-) membrane potential $V_m(x,t)$, defined as $V_m(x,t)=V_i(x,t)-V_o(x,t)$. At rest, before stimulation, the trans-membrane potential is not zero, according to Hodgkin-Huxley. It is equal to the resting potential $V_{i,R}$. There is an intracellular current $j_i(x,t)$ and a (trans-) membrane current $j_m(x,t)$; the latter is outgoing when positive. The model is defined by its physical constants: per unit length of the cable there is an axial resistance $r_i$ ($\Omega/cm$), a membrane resistance $r_m$ ($\Omega \cdot cm$) and a membrane capacitance $c_m$ (F/cm). The radius of the cable, which does not appear anywhere else, enters into $r_i$, $r_m$, and $c_m$. The latter parameters are often combined into the following characteristic constants: length constant $\lambda = \sqrt{r_m/r_i}$ and membrane time constant $\tau_m = r_m c_m$ [Jo 1995].

Here is the resulting antenna-version of the cable equation (derivation below):

$$c_m \frac{\partial V_i(x,t)}{\partial t} = \frac{1}{r_i}\frac{\partial^2 V_i(x,t)}{\partial x^2} - \frac{1}{r_m}\left[V_i(x,t) - V_{i,R}\right] + \left\{\frac{1}{r_m}V_0(x,t) + c_m \frac{\partial V_0(x,t)}{\partial t}\right\}. \qquad (3.1)$$

It is a second order partial differential equation with a boundary condition in space *x* and an initial condition in time *t*. The boundary condition in *x* may be *open* or *closed*. In case of the *open* boundary condition the cylinder represents a segment of a longer cell (axon or dendrite). In case of the *closed* boundary condition the cylinder may be considered to represent a crude model of a bipolar cell. The antenna version of the cable equation differs from the normal version by the presence of the term in curly brackets. This term is the driving term that carries the antenna signal.

In order to understand the antenna-version of the cable equation we now take a look at its derivation.

The current $j_i(x,t)$ is driven, according to Ohm, by the negative gradient of the potential $V_i(x,t)$,

$$r_i \cdot j_i(x,t) = -\frac{\partial V_i(x,t)}{\partial x}. \qquad (3.2)$$

A membrane current $j_m(x,t)$ is "leaking" through the membrane and decreases $j_i(x.t)$ when positive, or increases $j_i(x.t)$ when negative.

$$\frac{\partial j_i(x,t)}{\partial x} = -j_m(x,t). \tag{3.3}$$

The membrane current $j_m(x,t)$ consists of two parts: an Ohmian part and a capacitive part,

$$j_m(x,t) = \frac{V_i(x,t) - V_{i,R} - V_o(x,t)}{r_m} + c_m \frac{\partial(V_i(x,t) - V_{i,R} - V_o(x,t))}{\partial t}. \tag{3.4}$$

We take the partial derivative of eq. (3.2) with respect to $x$ and replace $\frac{\partial j_i(x,t)}{\partial x}$ by using (3.3). The term $j_m(x.t)$ is then replaced by the right hand side of eq. (3.4). Collecting terms we finally get the antenna-version of the cable equation (3.1) above.

## 3.2. Solution method

There are very advanced program packages available to solve the cable equation, without and with extra-cellular potential. These are NEURON, GENESIS, NEST, XPAUT, just to name a few. They are described e.g. by Brette et al. [Br 2007]. For simplicity purposes we wanted to have an easy way to insert the antenna signal $V_o(x,t)$ created by the extra-cellular stimulation current $j_o(x,t)$ in a given target volume. For that reason we wrote a FORTRAN program that does exactly that.

The boundary conditions are chosen as follows:

- Closed-end boundary conditions at $x=0$ and $x=l$, which means zero electric field at both ends of the cable; as has been said before, the cylinder thus becomes a crude model for a bipolar cell.
- $V_o(x,t)=0$ and $V_i(x,t)=V_{i,R}$ for $t=0$ and all values of $x$. This means that the extra-cellular potential is equal to zero and the intracellular potential is equal to the rest potential, at the beginning of stimulation.

We use the Runge-Kutta-4 method to integrate in t-direction, starting at t=0, and we use the Numerov method for integration in x-direction [Sc 1990]. At each step of the Runge-Kutta integration we employ a Liebmann relaxation with as many repetitions as needed for the information, coming from the boundaries in x, to migrate to all points of the x-mesh [Sc 1990]. For the Runge-Kutta method we use a variable mesh width, adapted to the time variation of the driving function $V_o(x,t)$.

After the main code has found a solution, this solution is inserted into the original equation in an independent subroutine. Only errors in the range of errors for numerical differentiation are accepted. Bad convergence, eventually coming from an unrealistic choice of cell parameters, will be detected in this way.

For the purpose of checking our results we have written a second FORTRAN code using a different numerical method: The unknown solution $V_i(x.t)$ is expanded into a Fourier series of cosine functions in $x$. The advantage of this method is that every term of the series automatically satisfies our boundary condition in $x$. The unknown functions to be determined are the expansion coefficients of the series, as functions of $t$; the Fourier series is cut when 10-digit stability of the solution is obtained [Sc 1990]. The results of this second computer code are found to be identical to the ones of the first code.

## 3.3. Examples

Let us see what happens when we put our model nerve cylinder into the retina, as shown in Fig. 3.2, and turn on a stimulation signal that travels in extracellular space. Since the cylinder is electrically closed at both ends by the chosen boundary condition it cannot be a section of a dendrite. But we may consider it a very crude model of a bipolar cell.

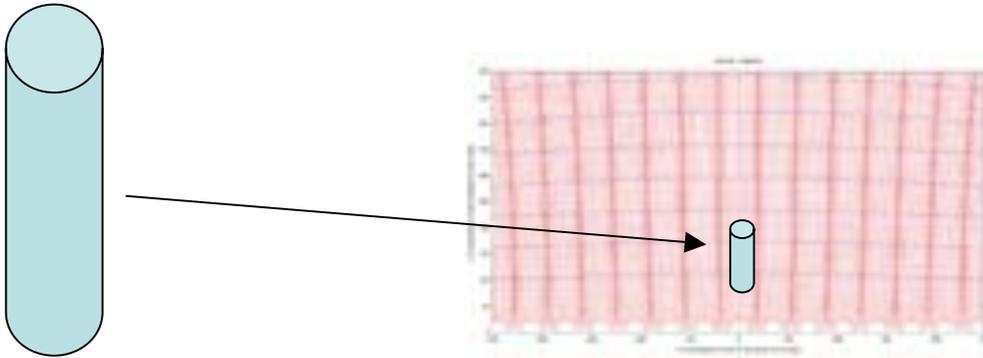

Fig. 3.2. The probe cylinder inserted into the electric stimulation field in the retina.

We choose parameters needed to solve the cable equation as follows : (1) length of the cylinder $l=100$ $\mu m$, (2) electric field strength $E=1$ millivolt/micrometer in the middle of the target volume, (3) several choices of cell constants, such as length constant $\lambda = 50$ $\mu m$, membrane time constant $\tau_m = 1$ ms, (4) parameters in the mathematical function that describes the time profile of the stimulation signal.

The electric field around the probe cylinder is taken from field calculations as shown in Sect. 2. In the present first calculation we let the electric field have a constant strength across the length of the cylinder, i.e. a linear increase (or decrease) of the extracellular potential is assumed; a potential that varies with x by a power law with any integer or non-integer power of x would also be accepted by the code.

We also need the time profile of the stimulating extracellular potential. In the present example we choose a monophasic rectangular stimulation potential with a length of $T=500$ $\mu s$, as shown in Fig. 3.3a. The resulting extracellular current is approximated by a superposition of two exponential functions, and is shown in Fig. 3.3b. It is this function that enters into the code.

The code outputs the solution $V_m(x,t) = V_i(x,t)-V_o(x,t)$, which is a function of $x$ and $t$. First let us consider the x-dependence, for some distinct values of $t$. For $t=0$ we have the chosen resting potential $V_m(x,t) = -0.070$ V for all values of $x$ (not plotted). At $t = 5$ $\mu s$ a small deviation from the resting potential appears at both ends of the cylinder. For a cathodic pulse, the end of the cylinder that is closer to the electrode ($x \approx 0$ $\mu m$) shows a small hyperpolarization while the end away from the electrode ($x \approx 100$ $\mu m$) shows a small depolarization. The depolarization/hyperpolarization increases with time, from $5$ $\mu s$ over $10$ $\mu s$, $20$ $\mu s$ to $50$ $\mu s$. Due to the fact that we have chosen a constant extracellular electric field across the length of the cylinder the solution is skew-symmetric with respect to $x=50$ $\mu m$; in more realistic calculations the extracellular field is taken from model calculations akin to those described in Sect. 2.

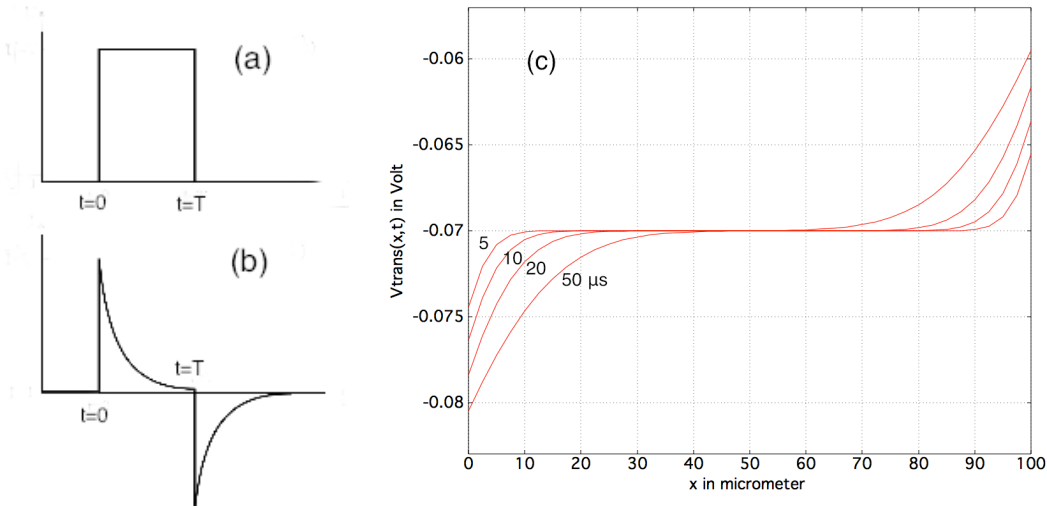

Fig. 3.3. Stimulation signal entering the probe cylinder. (a) the cathodic voltage pulse of - 2 V and duration $T=500$ μs applied to the electrode, (b) the time profile of the resulting current (in arbitrary units), (c) the transmembrane potential in the cylinder after 5 μs, 10 μs, 20 μs and 50 μs; the resting potential is set to -0.07 V.

We see that the probe cylinder picks up the stimulation signal in a way a rod antenna would pick up an electromagnetic signal. There is, however, an important difference: a rod antenna is designed to operate as a resonance antenna while the probe cylinder does not show a resonant behavior. We do not expect the passive cable equation to yield resonances. This might change when we go from the passive cable equation to an active one.

Next, we want to see how the depolarization at a given value of $x$ varies with time $t$. We choose $x=95$ μm, a value that corresponds to presynaptic space in our crude model of a bipolar cell. For the stimulation signal shown in Fig. 3.3a,b we see the result in Fig. 3.4. With a lag of time that depends on the chosen cell parameters, the transmembrane potential $V_m(x,t)$ follows the stimulating potential. Since the stimulating current is biphasic, see Fig. 3.3b, a hyperpolarization follows the depolarization. The intracellular response to the signal is of *yes/no*-type.

Fig. 3.4. For the stimulation signal shown in Fig. 3.3a,b the trans-membrane potential is plotted as a function of time. The value of $x$ is kept fixed at $x=95$ μm.

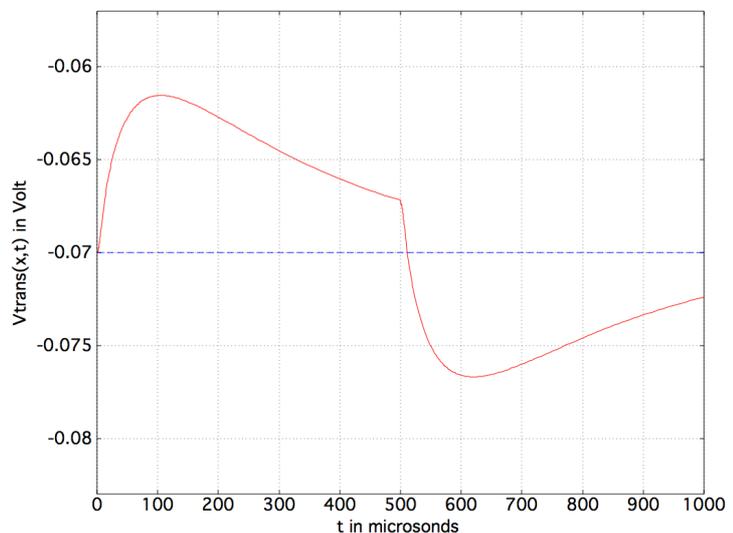

If we want depolarization only, we have to break the current (and discharge the electrode later with a sub-threshold current), as proposed in an earlier paper [Sc 2007]. Fig. 3.5 shows an example. The stimulation electrode is disconnected at $\Theta=150$ $\mu s$. The depolarization is slowly decaying during latency, $t >150$ $\mu s$. The signal is of *yes/latency* type.

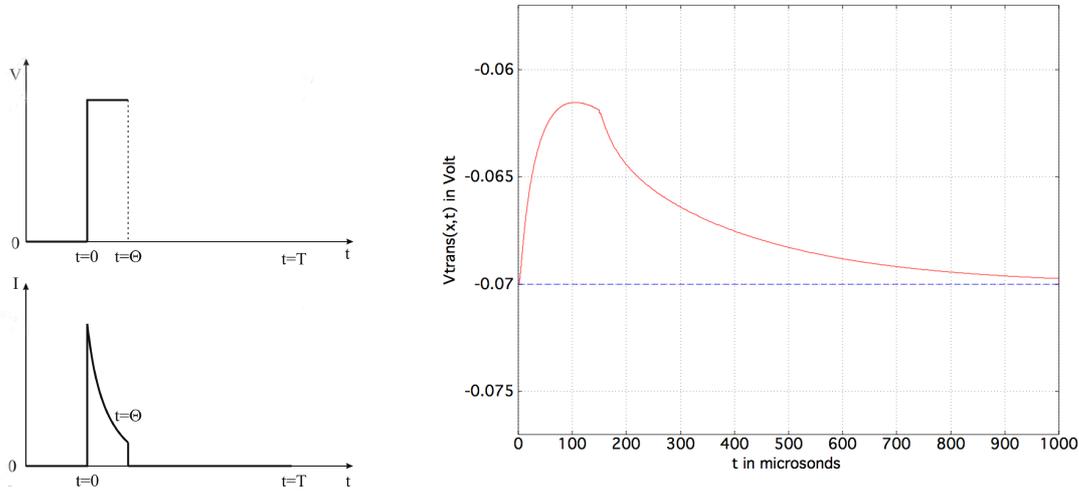

Fig. 3.5. The electrode is disconnected at $t=\Theta=150$ $\mu s$ and kept floating during latency $\Theta<t<T$ (it will be discharged at a later time not shown).

An interesting application of the computer code is seen in Fig. 3.6. Here we start from an example similar to the one of Fig. 3.5. The value of $\Theta$, however, is set to *500 $\mu s$*. At this value of $t$ the electrode has injected the same amount of charge into the retina as it did in the example of Fig. 3.3, with the difference that a latency period now follows the first part of the current pulse. We are interested in the following question: What do we get if we keep the injected charge fixed, and make the current pulse stronger and shorter? For simplicity we double the current strength without altering its time profile. Since we have approximated the time profile by two exponential functions it is easy to calculate a value for $\Theta$ such that the injected charge is the same for a current twice as strong. We obtained $\Theta=76$ $\mu s$. Fig. 3.6 shows the result of the calculation. The full red line is the result for $\Theta=500$ $\mu s$; the little kink at $t=500$ $\mu s$ can hardly be seen because, at this time, the current has become very small. The broken blue line shows the result for $\Theta=76$ $\mu s$ and the same amount of injected charge. The result is rather amazing. The shorter pulse yields about twice as much depolarization, and the area under the blue curve is not smaller than the area under the red one.

Fig. 3.6. The transmembrane potential $V_m(x,t)$, at $x=95\mu m$, is plotted for the *same* injected charge but two different pulse lengths. In case of the full red line the pulse length is $\Theta=500$ $\mu s$ plus latency. In case of the broken blue line the pulse is much shorter, namely $\Theta=76$ $\mu s$ plus latency; the pulse length has become shorter by doubling the strength of the current.

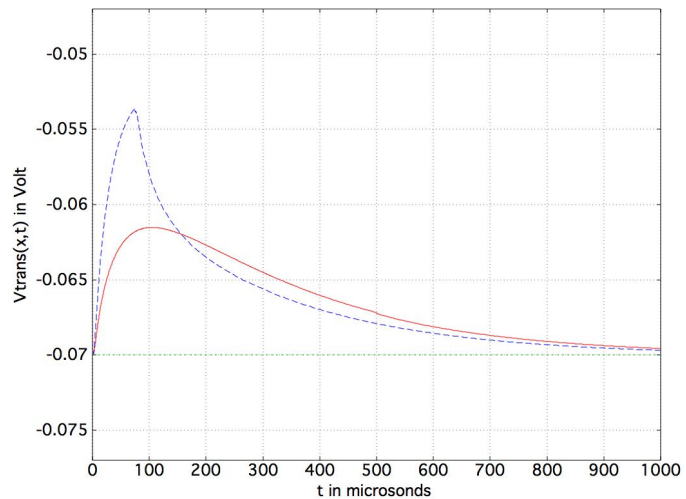

Does this result give us the key to sequential stimulation? As has been discussed in Sect. 2, the difficulty of sequential stimulation lies in the fact that we need a tremendous number of time slices in order to avoid time overlap of current pulses. As compared to the example of Fig. 3.4 we are already gaining a factor of 10 in the length of time slices, because the process of slowly discharging electrodes via a resistor does not lead to cross talk. Of course, this result is a theoretical modeling result and has to be experimentally tested.

## 4. Summary and conclusion

We presented and discussed two computational models to be used as tools in experimental research in retinal implants.

In the first model, electric field or current lines, as well as equipotential lines are calculated. In the shown examples the field is produced in the retina by multi-electrode arrays in subretinal position. Multi-electrode arrays in epiretinal position can be treated equally well: only the plotting routine has to take care of the different direction (*outward* instead of *inward*) of the *z*-axis.

In the second model, a target volume in the retina is chosen, and the passive Heaviside cable equation is solved inside the target volume in order to obtain information on the depolarization of the cell membrane.

In both models, a continuum approximation is assumed, i.e., the cell structure is smeared out except for a cylindrical section of a process of a neuron, which serves as a probe in the second model. The qualitative and quantitative usefulness of the computational tools is demonstrated in several examples.

The monopole activation of one single electrode corresponds to viewing a small bright object in a dark room. Despite its simplicity this case gives rise to the following question: How small may an electrode be? The field strength decreases rapidly with distance from the electrode because the current can spread out into the full solid angle of a hemisphere. To obtain the needed penetration depth of the stimulation signal we may need relatively large electrodes, or we need some bunching of current lines due to the presence of neighboring activated electrodes. Our calculation shows that already a second electrode that forms a dipole together with the first one yields some bunching effect.

An extreme case of bunching is produced by cross talk. If we reverse the above scenario of a small white object in a dark room by presenting a white wall with a small dark object in the middle, the current lines become almost parallel being bunched together by the current lines of many other electrodes. Things become even worse when the vitreous has been replaced by silicone oil because, in this case there is even less space for the current to spread out. A stagnation point will appear in (or near) the middle of the electrode array and might be mistaken for a dark object.

In principle, cross talk can be avoided by sequential stimulation as shown. The trouble is that many time slices are needed to avoid time overlap of activation. Model calculations show that time overlap of 24 electrodes, already, can ruin an image. With 9 simultaneously activated electrodes (8 cross-talking electrodes around a center electrode) imaging might still be possible in principle, unless one wants to do the fine-tuning needed for transmitting a grey scale, such as the one that helps in Fig. 1.1 to recognize the face.

Another way to avoid the cross-talk problem is stimulation by electrodes that form multipoles. Problems are also encountered here. When cathodes and anodes are close together, part of the current takes the shortest way and travels from electrode to electrode without entering into the retina. This part of the current is lost for stimulation. The loss becomes bigger when there is an adhesive layer of saline between the chip and the retina, because the electric conductivity of a saline is much higher than the average conductivity of a tissue. Two calculations have been performed to illustrate this effect. More calculations will be needed solve this dilemma.

In conclusion we turn back to the Introduction. In order to achieve the desired image quality of Fig. 1.1 one definitely has to avoid cross talk and one also has to avoid stimulation in area C of Fig. 1.2. Sequential stimulation seems to be a solution. However, this stimulation modality has flaws, because the road to implants with many more pixels is closed because of the limited number of time slices and because stimulation in area C of Fig. 1.2 is not blocked. Another way out seems to be multipolar stimulation. It allows bunching of field lines, and the current does not reach area C of Fig. 1.2. These statements are based on theoretical models, and therefore have to be verified or falsified by experiment.